\documentclass{phyeauth}

\journal{Physica E}
\usepackage{amssymb}
\usepackage{graphicx}
\usepackage{latexsym}

\newcommand{\fig}[1]{Fig.~\ref{#1}}

\newcommand{\eq}[1]{Eq.~(\ref{#1})}

\newcommand{\bref}[1]{Ref.~\cite{#1}}
\newcommand{\brefs}[1]{Refs.~\cite{#1}}

\renewcommand{\d}{{\rm d}}

\newcommand{\ub}{{\rm ub}}
\newcommand{\bd}{{\rm b}}
\newcommand{\piad}{\pi_{\rm ad}}

\begin{document}

\begin{frontmatter}

\title{Movements of molecular motors: \\
  Ratchets, random walks and
 traffic phenomena} 

\author[mpi]{Stefan Klumpp \thanksref{thank1}},
\author[amst]{Theo M. Nieuwenhuizen} and
\author[mpi]{Reinhard Lipowsky}

\address[mpi]{Max-Planck-Institut f\"ur Kolloid- und
  Grenzfl\"achenforschung, 14424~Potsdam-Golm, Germany}
\address[amst]{Instituut voor Theoretische Fysica, Universiteit van 
  Amsterdam, Valckenierstraat~65, 1018 XE Amsterdam, The Netherlands}

\thanks[thank1]{
Corresponding author. 
E-mail: klumpp@mpikg-golm.mpg.de}

\begin{abstract}
  Processive molecular motors which drive the traffic of organelles in
  cells move in a directed way along cytoskeletal filaments. On large
  time scales, they perform motor walks, i.e., peculiar random walks 
  which arise from the repeated
  unbinding from and rebinding to filaments. Unbound motors perform
  Brownian motion in the surrounding fluid. In addition, the traffic of
  molecular motors exhibits many cooperative phenomena. In particular, it 
  faces similar problems as the traffic on streets
  such as the occurrence of traffic jams and the coordination of
  (two-way) traffic.  These issues are studied here theoretically using
  lattice models.
\end{abstract}

\end{frontmatter}

%%%%%%%%%%%%

\section{Introduction}

The idea of constructing nanometer-sized devices and machines has
created a lot of excitement during the last years. Despite the progress 
made, the functionality
of artificial nano-devices is, however, still rather limited. At the
same time, more and more biomolecular nano-machines have been
identified in the cells of living beings where they accomplish a huge
variety of tasks. Many of these molecular motors are now rather well
studied and were found to work with an amazing degree of
precision and efficiency as a result of
billions of years of evolution, \cite{Howard2001,Schliwa2003}.  In the
following, we will focus on one class of molecular motors which has
been studied quite extensively during the last decade, namely
processive cytoskeletal motors which drive the traffic of vesicles and
organelles within cells. These motors hydrolyze adenosinetriphosphate
(ATP) and convert the free energy from this chemical reaction into
directed movements along filaments of the cytoskeleton. This class of
motors contains kinesins and dyneins, which move along microtubules, and
certain myosins, which move along actin filaments. These motors walk
along the filaments by performing discrete steps with a step size
which corresponds to the repeat distance of the filament, 8~nm for
kinesins 
%and dyneins\footnote{The step size of dynein depends on load. At low load, 
%the step sizes are integer multiples of 8~nm \cite{Mallik__Gross2004}.} 
and 36~nm for myosin V.  They are called
processive if they make many steps while staying in contact with the
filament.

From a physical point of view, much of the interest in molecular
motors
is due to the fact that 
the difference in size compared to macroscopic engines implies also
conceptual differences.
%, despite many similarities such as the use of a
%chemical 'fuel' which provides the (free) energy converted into
%mechanical work. The difference in size is accompanied by a difference
%in the typical energy scale of the engine. 
The typical energy of
macroscopic motors is much larger than the thermal energy, $k_{\rm
  B}T$, while the typical energies of molecular motors are of the
order
of $k_{\rm B}T$. For example, the hydrolysis of ATP releases about 
20~$k_{\rm B}T$.
On the one hand, molecular motors have to cope with perturbations arising from
thermal fluctuations; on the other
hand, the unavoidable presence of noise suggests that evolution has
created motors which make use of this noise in order to generate
work or directed movement.

From a technological viewpoint, the amazing properties of single
biological motor molecules and the complexity of the systems into
which they are integrated in the cell provide inspiration for the
design of artificial nanoscale transport systems
%and suggest that
%integrating biological motor molecules into synthetic devices is a
%promising route towards a nanotechnology of molecular motors
\cite{Hess_Vogel2001,Boehm_Unger2004}.

In this article, we discuss several theoretical aspects of the motor
movements. 
In section
\ref{sec:Ratchets}, we start by summarizing some recent experimental
results and discuss the question whether noise-driven mechanisms are
used by these motors. In section \ref{sec:RW}, we discuss another
effect of noise, namely the detachment of motors from
their tracks due to thermal fluctuations, which leads to peculiar
random walks. We derive the asymptotic behavior of these random walks
using the statistical properties of the returns of motors to the
filament. Finally, in section \ref{sec:traffic}, we summarize our
recent studies of traffic problems which arise in systems with many
molecular motors due to their mutual exclusion from binding sites of
the filaments. These topics are also addressed in a recent longer 
review article \cite{Lipowsky_Klumpp2005}.

\section{Active movements of molecular motors}
\label{sec:Ratchets}

\begin{figure}[tb] 
  \begin{center} 
    \includegraphics[angle=0,width=.5\textwidth]{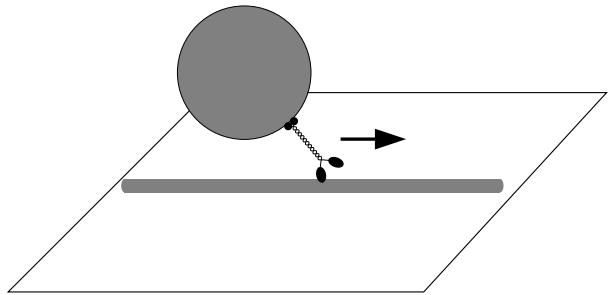}
    \caption{The 'bead assay' constitutes a biomimetic model system: A 
      molecular motor transports a (glass or latex) bead along a
      filament which is immobilized on a surface. }
    \label{fig:beadassay} 
  \end{center} 
\end{figure}

Molecular motors can be studied outside cells using biomimetic model
systems. In these experiments, the biological complexity is reduced to
a minimal number of components, namely motors, filaments, and ATP.  An
example is shown in \fig{fig:beadassay}.  By these experiments, one can
observe movements of single motor molecules and measure
transport properties such as velocities, step sizes, and forces
\cite{Howard2001,Schliwa2003}.
%This implies that,
%on the one hand, these experiments provide insight into the motor
%mechanisms and the dynamics of the single steps, and, on the other
%hand, they allow to measure systematically the transport properties as
%a function of certain external control parameters. In addition, they also 
%provide the basic setup for applications in nanotechnology, see 
%\bref{Hess_Vogel2001}.

On the one hand, these experiments provide insight into the motor
mechanisms. A major breakthrough was
%With respect to the motor mechanisms, a main breakthrough was 
to resolve the discrete steps of the motors and to measure the step
size of the motor which corresponds to the repeat distance of the
filament. This has first been achieved for kinesin using an optical
tweezers setup \cite{Svoboda__Block1993}. More recently, it has been
shown that kinesin \cite{Asbury__Block2003,Yildiz__Selvin2004} and
also myosin~V \cite{Yildiz__Selvin2003} move in a hand-over-hand way,
i.e., that the two heads of the dimer step forward in an alternating
fashion such that the rear head always moves in front of the leading
head, similar to human walking.  Further progress is expected from
combining mechanical methods and particle tracking with fluorescence
techniques.

On the other hand, using these biomimetic motility assays, one can measure
the transport properties systematically varying external control
parameters.
%With respect to the systematic measurement of the transport properties
%as functions of external control parameters, 
Here the main focus has been on the velocity as a function of the ATP
concentration and of the force applied with, e.g., optical tweezers to
oppose the movements, see, e.g., \bref{Visscher__Block1999}. Other
quantities that have been measured are the one-dimensional diffusion
coefficient of motors bound to filaments or the randomness parameter
and the walking distance before unbinding from the filament. These
measurements have stimulated a large amount of theoretical work, see,
e.g.\ \brefs{Juelicher__Prost1997,Lipowsky2000b,Astumian_Haenggi2002},
modeling the walks of motors along filaments in order either to fit
the experimental data or to find out the generic properties of these
walks. For example, it turns out that the motor velocity as a function
of the ATP concentration is given by a universal relationship which
should be valid for many types of motors
\cite{Lipowsky2000a,Lipowsky_Jaster2003}.

Since nanometer-sized molecular motors have to live in a noisy
environment, it has soon been speculated whether these motors exploit
the noise to generate their directed movements, and various variants
of ratchet models have been proposed as reviewed in
\brefs{Juelicher__Prost1997,Lipowsky2000b,Astumian_Haenggi2002}. In the
simplest case, the conformational changes associated with the chemical
cycle of the motor rectify the one-dimensional Brownian motion along
the filament. While such a simple mechanism is not consistent with the
measurements for dimeric motors such as conventional kinesin or myosin
V, it can describe movements of processive monomeric motors such as
the monomeric kinesin KIF1A
\cite{Okada_Hirokawa1999,Tomishige__Vale2002,Okada__Hirokawa2003} (for
other processive monomeric motors, see the review
\cite{Schliwa_Woehlke2003}). These monomeric motors exhibit biased,
but strongly diffusive movements along the filaments similar to what
one obtains in the simplest ratchet models. Interestingly, coupling
two such ratchets by a spring \cite{Ajdari1994,Klumpp__Wald2001} leads
to a driving mechanism which is independent of the diffusion along the
filament.  Similarly, dimerization of monomeric kinesin results in a
higher velocity and smaller diffusion coefficient of the
filament-bound motors \cite{Tomishige__Vale2002}.
%Many monomeric motors, however, do
%not remain bound to the filament during the whole chemical cycle.
These results suggest that the dimerization of motors has two effects:
It allows the motor to stay bound to the filament for many chemical
cycles -- in contrast to many (therefore unprocessive) monomeric motors
-- and in addition, it allows the motors to move by a more efficient
mechanism which, in contrast to the motility of monomers, does
not rely on diffusion along the filament.

\section{Random walks arising from many diffusional encounters with filaments}
\label{sec:RW}

\subsection{Motor walks in open compartments}

% Fig.2
\begin{figure}[tb] 
  \begin{center} 
    \includegraphics[angle=0,width=.5\textwidth]{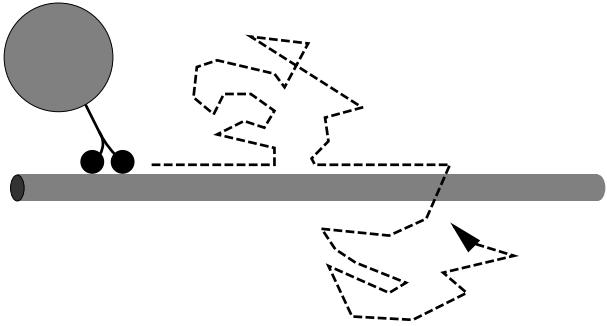}
    \caption{Random walk of a molecular motor: The motor performs directed  
      movement along a filament and unbinds from it after a 
      certain walking distance. The unbound motor diffuses in 
      the surrounding fluid until it rebinds to the filament and 
      resumes directed motion.} 
    \label{fig:randWalk} 
  \end{center} 
\end{figure}

The fact that nanometer-sized motor molecules work in a
noisy environment has another consequence: Even processive motors
do not move along a filament forever, but unbind from it after a
certain binding time (which corresponds to a typical walking
distance), because the binding energy is finite and can be overcome by
thermal fluctuations. For a single kinesin or myosin~V motor, the
binding times and walking distances are of the order of 1 s and 1
$\mu$m, respectively. Much longer walking distances can be obtained if
several motors form a complex or if a cargo is transported by a larger
number of motors. Note that the unbinding 
%which arises as the effect of unavoidable fluctuations 
may also have a biological function by allowing the motors to diffuse 
around obstacles %such as other proteins bound to 
on the filament.

Unbound motors perform simple Brownian motion until they rebind to the
same or another filament. 
On large time scales, the combination of
active directed movements along filaments and non-directed Brownian
motion leads to peculiar random walks, called motor walks in the 
following, which consist of alternating sequences of the two types of movements
as sketched in \fig{fig:randWalk} \cite{Ajdari1995,Lipowsky__Nieuwenhuizen2001}.

In order to determine the effective transport properties of these
motor walks, we have studied several simple arrangements of filaments
embedded in compartments of various geometries. A particularly simple
but intriguing case consists of a single filament and a set of
confining walls, which restrict the diffusion of unbound motors.  In
the simplest case, there are no confining walls and the unbound motors
can diffuse freely in the full three-dimensional space (similar
behavior is obtained for a half space geometry which is more easily
accessible to experiments). By placing the filament in a quasi
two-dimensional slab or in a cylindrical tube (geometries, which are
also accessible to {\it in vitro} experiments), diffusion can be
restricted along one or two dimensions perpendicular to the filament.

We have studied the motor walks by mapping them to random walks on
a lattice \cite{Lipowsky__Nieuwenhuizen2001}. A line of lattice sites
represents the filament. Motors at these sites perform a biased random
walk and move predominantly into one direction, which we choose to be
the positive $x$ direction. In a discrete-time description, they move with a small probability $\epsilon/2d$ to each of the adjacent non-filament sites and thus unbind
from the filament. At the non-filament sites the motors perform simple
symmetric random walks and move to each neighbor site with probability
$1/2d$ ($d$ denotes the spatial dimension) and rebind to the filament
with probability $\piad$ when they reach again a filament site.
Confining walls are implemented as repulsive boundaries, at which all
attempted movements towards the walls are rejected.

\begin{figure}[tb] 
  \begin{center} 
    \includegraphics[angle=-90,width=\columnwidth]{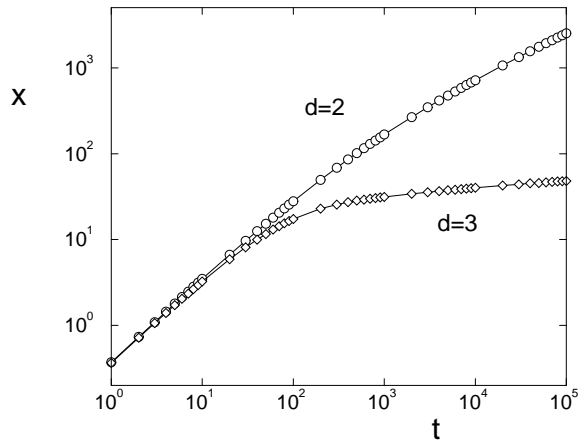}
    \caption{Average displacement $x$ as a function of time $t$ for the 
      motor walks as shown in \fig{fig:randWalk} in two- 
      and three-dimensional systems without confining walls. The 
      displacement grows 
      as $x\sim t^{1/2}$ and $x\sim \ln t$ for $d=2$ and $d=3$, 
      respectively. } 
    \label{fig:displacement} 
  \end{center} 
\end{figure}

We have used scaling arguments, computer simulations, and exact
solutions of the master equations to study the drift and diffusion 
behavior arising from the motor walks
\cite{Lipowsky__Nieuwenhuizen2001,Nieuwenhuizen__Lipowsky2002,Nieuwenhuizen__Lipowsky2004,ProcTaiwan}.
The motor walks exhibit anomalous drift behavior and strongly enhanced
diffusion parallel to the filament due to the repeated binding and
unbinding. At large times, motors move with an effective velocity
given by $v_\bd P_\bd$, where $v_\bd$ is the velocity of the bound motor
and $P_\bd$ is the probability that
the motor is bound to the filament. In the tube geometry, $P_\bd$ is
time-independent at large times and given by the equilibrium of
binding/unbinding and diffusion perpendicular to the filament. In the
slab and half-space geometries as well as in two- and
three-dimensional systems without confining walls, a steady state cannot be
reached at any time, because motors can rebind to the filament after
arbitrarily large excursions, and the typical size of these
%longer one observes the
%motors, the larger the 
excursions, that contribute to the average
behavior, increases with time $t$. Therefore, $P_\bd$ and the effective velocity are
time-dependent in these cases, namely $P_\bd(t)\sim t^{-d_\perp/2}$
for compartments with $d_\perp$ dimensions of unconfined diffusion
(for $d$-dimensional systems without confining walls, we have
$d_\perp=d-1$). The time-dependent effective velocity implies that the
average displacement of the motors grows sublinearly. In the
effectively two-dimensional slab geometry (as well as on a
two-dimensional lattice without confining walls), the displacement
behaves as $x(t)\sim\sqrt{t}$ at large times, and in the half space
(or full three-dimensional space), it is given by $x(t)\sim \ln t$, see 
\fig{fig:displacement}.

In the following, we will give a simplified description which explains
these features and relates them to known results from the theory of
random walks.

\subsection{Asymptotics and return to the filament}
\label{sec:asymptotics}

For simplicity, we discuss the case of a single infinitely long
filament embedded into $d$-dimensional space with $d=2$ or $d=3$.
Motor particles binding to this filament walk along it with velocity
$v_\bd$ until they unbind. Unbinding occurs with a rate
$\sim\epsilon$, so that the motors perform straight movements over the walking distance 
%$x_s$ given by the walking distance, 
$\Delta x_\bd\sim v_\bd/\epsilon$.  The
diffusive excursions between two bound walks bring the motors back to
the filament. (For simplicity, we take the sticking probability
$\piad$ to be one. If $\piad<1$, the motor has to return to
the filament $1/\piad$ times before rebinding.) The distribution
$\psi(\tau)$ of the excursion times $\tau$ is therefore given by the
distribution of the return times of a random walker's return to a line
in $d$ dimensions or, if we consider the projection into the plane
perpendicular to the filament, of the return to the origin in $d-1$
dimensions. This is a classical problem in the theory of random walks
which was solved by Polya in 1921 and has lead to the remarkable
result that the return of a random walker to the origin is certain on
one- and two-dimensional lattices, but not in three dimensions
\cite{Polya1921}. For the molecular motors, this implies that they
will return to the filament with certainty.

If we are only interested in the movement parallel to the filament, we
can consider the excursions away from the filament as periods of rest.
In addition, for large times, the duration of an excursion is typically 
much longer 
than the time the motor is bound to the filament; therefore 
we can consider the walks along the filament as effectively instantaneous 
steps of size $x_s\sim\Delta x_\bd\sim v_\bd/\epsilon$ (If $\piad<1$, the 
effective step size is $x_s\sim v_\bd\piad/\epsilon$.
The motor walks are then described by continuous time random walks
with the dwell time distribution $\psi(\tau)$ given by the
distribution of the excursion times. For this type of random walks,
solutions can be obtained using Fourier--Laplace transforms
\cite{Weiss1994}. A short summary of this method is given in
appendix \ref{sec:app}. In that way, one finds that the Laplace transforms of the
first two moments of the position of this random walker are given by
%\begin{equation}\label{asMoments_PSIabh}
%  x(s)=\frac{\langle x_s \rangle\psi(s)}{s[1-\psi(s)]}\qquad {\rm and}\qquad x^2(s)=\frac{\langle x_s\rangle^2 \psi(s)}{s[1-\psi(s)]}+\frac{2\langle x_s^2\rangle \psi^2(s)}{s[1-\psi(s)]^2}
%\end{equation}
\begin{equation}\label{asMoments_PSIabh}
  x(s)=\frac{\langle x_s \rangle\psi(s)}{s[1-\psi(s)]}
\end{equation}
and
\begin{equation}\label{asMoments_PSIabh2}
x^2(s)=\frac{\langle x_s\rangle^2 \psi(s)}{s[1-\psi(s)]}+\frac{2\langle x_s^2\rangle \psi^2(s)}{s[1-\psi(s)]^2}
\end{equation}
where $\psi(s)$ is the Laplace transform of the waiting time
distribution $\psi(s)\equiv\int_{0}^{\infty} \d t\, \psi(t) e^{-st}$.
In general, $\langle x_s\rangle$ and $\langle x_s^2\rangle$ are the 
moments of the step size distribution; for the motor walks, they are 
given by the active walks along the filament and we take them to be given 
by $x_s=v_\bd/\epsilon$ and $x_s^2$, respectively.
From these relations, the average displacement and the dispersion of
the motor walks for large times can be obtained by inverting the
Laplace transform. In particular, these relations imply that the
asymptotic displacement of the motors is given by the large-time (or
small $s$) behavior of the distribution of return times to the origin
in $d-1$ dimensions.

Normal drift behavior with $x(t)\sim t$ is obtained, as long as
$\psi(\tau)$ has a finite mean value, $\bar\tau$, and thus
$\psi(s)\approx 1-\bar\tau s$ for small $s$.  If, however,
$\psi(\tau)$ decays slower than $\sim\tau^{-2}$ at large $\tau$, the
mean dwell time diverges (which implies a divergence of
$[\psi(s)-1]/s$ for small $s$), and anomalous drift is obtained.
%, i.e., the average displacement $x(t)$ grows slower than linearly with time.
The latter behavior occurs in our case (where the waiting times are
given by return times to the filament) if the diffusion away from the
filament is not restricted.

In the two-dimensional case, the return time distribution behaves as
$\psi(\tau)\approx 1/(2\sqrt{\pi} \tau^{3/2})$ for large $\tau$ or
$\psi(s)\approx 1-\sqrt{s}$ for small $s$ as shown in 
appendix \ref{sec:app:return}.  Inserting this into
\eq{asMoments_PSIabh} and inverting the Laplace transform, we obtain
%\begin{equation}
%  x(s)\approx \frac{x_s}{s^{3/2}}=\frac{v_\bd}{\epsilon s^{3/2}} \qquad{\rm and}\qquad  x(t)\approx \frac{2 x_s\sqrt{t}}{\sqrt{\pi}}=\frac{2v_\bd\sqrt{t}}{\epsilon\sqrt{\pi}}
%\end{equation}
%
\begin{equation}
  x(s)\approx \frac{x_s}{s^{3/2}}=\frac{v_\bd}{\epsilon s^{3/2}}
\end{equation}
%$x(s)\approx x_s s^{-3/2}$ 
and
\begin{equation}
 x(t)\approx \frac{2 x_s\sqrt{t}}{\sqrt{\pi}}=\frac{2v_\bd\sqrt{t}}{\epsilon\sqrt{\pi}}
\end{equation}
for small $s$ and large $t$, respectively.
%for large $t$.  
Similarly, in the three-dimensional case, the return time distribution
is $\psi(\tau)\approx 2\pi/(3 \tau\ln^2 \tau)$ for large $\tau$ or
$\psi(s)\approx 1-2\pi/(3\ln s^{-1})$ for small $s$, see again appendix 
\ref{sec:app:return}, which leads to
\begin{equation}
  %x(s)\approx \frac{3 x_s \ln s^{-1}}{2\pi\, s}=\frac{3v_\bd \ln s^{-1}}{2\pi\epsilon\, s} \qquad{\rm and}\qquad 
x(t)\approx \frac{3 x_s}{2\pi}\ln t=\frac{3v_\bd}{2\pi\epsilon}\ln t.
\end{equation}

Likewise, we can obtain the dispersion $\Delta x^2(t)$ of the motors
from the second moment of the distribution arising from the encounters
with filaments. \eq{asMoments_PSIabh2} leads to $\Delta x^2(t)\approx
\frac{2(\pi-2)}{\pi}(v_\bd/\epsilon)^2 t+\frac{1}{2}t$ and $\Delta
x^2(t)\approx \frac{9}{4\pi^2}(v_\bd/\epsilon)^2 \ln t +\frac{1}{3}t$
in two and three dimensions, respectively, where we have added the
contribution due to the diffusion of unbound motors parallel to the
filament. Note that the broadening of the distribution of motors due
to the encounters with the filament is characterized by an anomalously
high effective diffusion coefficient of the order of
$(v_\bd/\epsilon)^2$ in two dimensions, while in three dimensions, the
leading term is the unbound diffusion, but with a large logarithmic
correction, again of the order $(v_\bd/\epsilon)^2$.  These results
agree with the corresponding asymptotic results from the exact
solution of the full master equations
\cite{Nieuwenhuizen__Lipowsky2002,Nieuwenhuizen__Lipowsky2004}.

\section{Traffic phenomena in many-motor systems}
\label{sec:traffic}

\subsection{Traffic jams and density patterns}

Finally, we consider systems with many interacting molecular motors.
The simplest type of motor--motor interaction is simple exclusion
which arises from the fact that a motor occupies a certain volume and,
in particular, excludes other motors from the binding site of the
filament to which it is bound as observed in decoration experiments,
see, e.g., \bref{Song_Mandelkow1993}. Exclusion is most important if motors
accumulate in certain regions along the filaments, where it leads to
the formation of molecular traffic jams.  These interactions are
easily incorporated into our lattice model by rejecting all movements
to occupied lattice sites 
\cite{Lipowsky__Nieuwenhuizen2001,Klumpp_Lipowsky2003,Klumpp__Lipowsky2005}. 
These models then represent new variants of
exclusion processes or driven lattice gas models, where the active or
driven movements are localized to the filaments.  In driven exclusion
processes, the state of the system depends crucially on the boundary
conditions, since the boundaries determine the current through the
system.

We have studied tube systems with a single filament located along the
axis of a cylindrical tube and with several types of boundary
conditions at the left and right tube ends
\cite{Lipowsky__Nieuwenhuizen2001,Klumpp_Lipowsky2003}. We use the
convention that the active movement along the filament is biased to
the right.  There are several possibilities to build such tube systems
artificially such as micropipette glass tubes or liquid microchannels,
but there are also several tubular compartments within cells for which
these tube systems provide simple descriptions. The most prominent
example for the latter is the axon of a nerve cell; another example is
provided by the hyphae of fungi.

The simplest situation is given by {\it periodic boundary conditions}
which can be solved exactly \cite{Klumpp_Lipowsky2003}. In this case,
motor particles reaching the right end of the tube simply restart
their movements at the left end. This leads to constant density
profiles for both the bound and unbound motors in the stationary
state. Diffusive currents in the radial direction vanish and binding
to the filament is locally balanced by unbinding.
%In this situation,
%which we call radial equilibrium, the bound and unbound motors
%densities $\rho_\bd$ and $\rho_\ub$, respectively, fulfill the
%condition
%\begin{equation}
%  \epsilon\rho_\bd(1-\rho_\ub)=\piad\rho_\ub(1-\rho_\bd).
%\end{equation}
The current of motors through the tube is given by
$J=v_\bd\rho_\bd(1-\rho_\bd)$ and the value of the bound density
$\rho_\bd$ is determined by the total number of motor particles in the
tube. If the number of motors within the tube is increased beyond an
optimal number (where $\rho_\bd=1/2$ and $J=v_\bd/4$), the current
through the tube decreases due to jamming of the motors.

\begin{figure}[tb] 
  \begin{center} 
    \includegraphics[angle=0,width=\columnwidth]{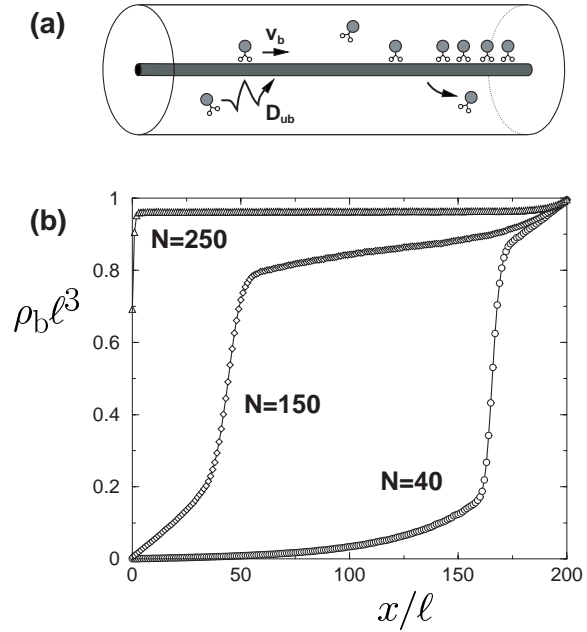}
    \caption{(a) Motors moving in a closed tube system which contains one 
      filament. The motors move with velocity $v_\bd$ along the
      filament to the right and build up a motor traffic jam at the
      right end of the system. In the stationary state, the bound
      motor current is balanced by a current of unbound motors to the
      left arising from the diffusive movement of the unbound motors
      which is characterized by the diffusion coefficient $D_\ub$.
      (b) Profiles of the bound motor density as a function of the
      coordinate $x$ along the filament for various total numbers $N$
      of motors within the tube. The jammed region becomes longer with
      increasing $N$.}
    \label{fig:tube} 
  \end{center} 
\end{figure}

In a {\it closed tube}, motors accumulate in front of the right end,
and a diffusive current of unbound motors to the left balances the
current along the filament in the stationary state
\cite{Lipowsky__Nieuwenhuizen2001,Klumpp__Lipowsky2005}. If the number 
of motors in the
tube is small, the motors are essentially localized at the right tube
end. Upon increasing the number of motors within the tube, a jammed
region at the right tube end builds up, separated from a low density
region to its left by a rather sharp interface, which provides
probably the simplest example for active pattern formation by
molecular motors.  The crowded domain spreads to the left at higher
motor concentrations until
%, and, eventually, if there are many motors in the tube, 
the filament is uniformly covered by motors and rather crowded, see
\fig{fig:tube}. Such density profiles have recently been observed for
a kinesin-like motor in fungal hyphae \cite{Konzack__Fischer2004}.
Let us note that these density patterns exist due to continuous 
consumption of ATP. If the 
ATP concentration is not kept constant within the closed compartment, 
then after burning all ATP, the 
density pattern will finally become homogeneous.

\begin{figure}[tb] 
  \begin{center} 
    \includegraphics[angle=-90,width=\columnwidth]{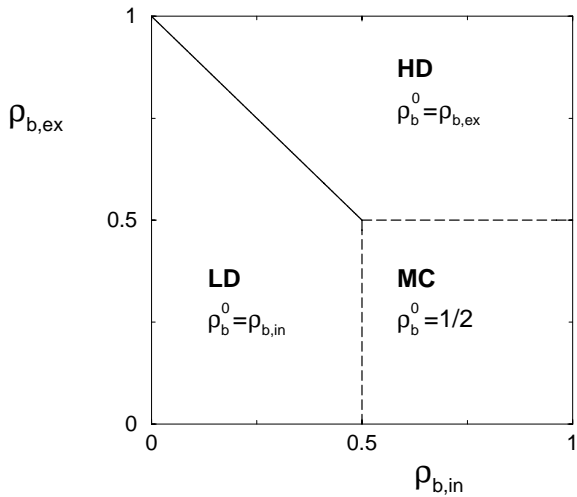}
    \caption{Phase diagram for the motor traffic through an open tube 
      which is coupled to motor reservoirs at the left and right end, 
      characterized by bound motor densities $\rho_{\rm b,in}$ and 
      $\rho_{\rm b,ex}$, respectively \cite{Klumpp_Lipowsky2003}. Three 
      phases -- a low density (LD) phase, a high density (HD) phase, and 
      a maximal current (MC) phase -- are distinguished by the value of 
      the bulk density $\rho_\bd^{0}$ in the interior of the tube. In 
      the case shown here, bound and unbound motor densities satisfy radial 
      equilibrium at the boundaries, for other choices of the boundary 
      conditions, the transition lines are shifted. }
    \label{fig:PD_OT} 
  \end{center} 
\end{figure}

Tubes with {\it open boundaries} which are coupled to reservoirs of
motors at both ends, so that motors enter the tube at the left end and
leave it at the right end, exhibit boundary-induced phase transitions
\cite{Klumpp_Lipowsky2003,Klumpp_Lipowsky_PRE2004}. There are three phases which are
distinguished by the 'bottleneck' which determines the motor current
through the tube. This 'bottleneck' can be the left boundary, the right
boundary or the interior of the tube. These three cases corresponds to
the low-density (LD), high-density (HD) and maximal-current (MC)
phase as shown in \fig{fig:PD_OT} . If changing the motor densities in the reservoirs at the
boundaries leads to a change in the 'bottleneck' position, a phase
transition occurs which can be either discontinuous (LD--HD) or
continuous (LD--MC and HD--MC). These types of phases and transitions
are known from the one-dimensional asymmetric simple exclusion process
(ASEP) \cite{Krug1991,Kolomeisky__Straley1998} which corresponds to
the dynamics along the filament in our model without the binding and
unbinding processes. The presence of the unbound state of the motors,
however, increases the number of possible boundary conditions, 
and since the phase transitions are
boundary-induced, the location of the transition lines within the
phase diagram is quite sensitive to that choice.
%, which may be hard to
%control in an experimental system, so that precise predictions will be
%difficult. 
At present, it is difficult to see which of these boundary conditions 
will be the simplest to implement experimentally.
Nevertheless, systems of molecular motors are promising
candidates for the experimental observation of boundary-induced phase 
transitions.

\subsection{Phase transitions in two-way traffic}

\begin{figure}[tb] 
  \begin{center} 
    \includegraphics[angle=-90,width=\columnwidth]{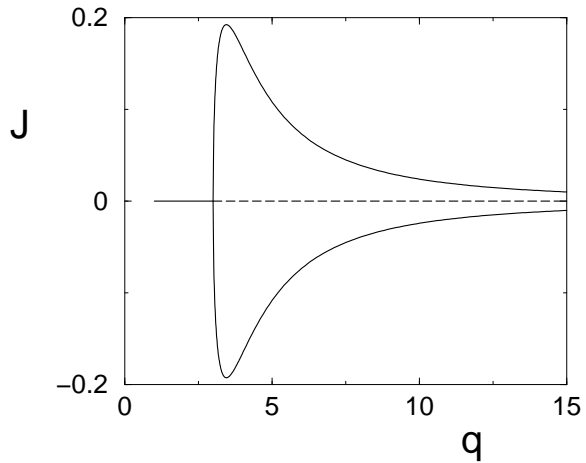}
    \caption{Spontaneous symmetry breaking in systems with two motor species moving into opposite directions along the same filament: The total motor current $J$ is zero for weak interaction $q<q_c$, where the filament is equally populated by both motor species, but non-zero for sufficiently strong motor--motor interactions with $q>q_c$, where one motor species is essentially excluded from the filament \cite{Klumpp_Lipowsky2004}. For very strong interaction, the current decreases because the filament becomes more and more crowded.}
    \label{fig:2spec} 
  \end{center} 
\end{figure} 

Another type of phase transitions occurs in systems with two-way
traffic of motors: While each motor moves either towards the plus-
or towards the minus-end of the corresponding filament, 
different types of motors move into opposite directions.  
%along the same filament. 
%Decoration experiments indicate that these motors interact
%in such a way that a motor is more likely to bind to the filament if
%another motor which belongs to the same species is already bound
%there, and that motors have a tendency to unbind more easily if they
%encounter a motor moving into opposite direction. This interaction is
%probably mediated by conformational changes or deformations of the
%filament upon binding of a motor. 
We have studied systems with two
species of motors moving into opposite directions along the same
filament \cite{Klumpp_Lipowsky2004}. The interactions of these motors 
are described by a
%and modeled the interactions of the motors using a model with a
single interaction parameter $q$ by which the binding and unbinding
rates are increased or reduced in order to enhance binding and reduce
unbinding if a motor of the same type is present at a neighbor site
and to enhance unbinding and reduce binding if a motor with opposite
directionality is bound at a neighbor site.
This type of interaction is suggested by decoration experiments.
%It turns out that if there are no such effectively attractive and
%repulsive interactions and motors interact only through their mutual
%exclusion, the motor current will be very small because motors often
%block each other and continue moving only after one of them detached
%from the filament. 
For equal concentrations of
both motor species and for sufficiently strong interaction with $q>q_c$
(where $q_c$ is the critical value of the interaction parameter which
depends on the overall motor concentration),  there is spontaneous 
symmetry breaking,
so that one motor species occupies the filament, while the other one
is largely excluded from it, see \fig{fig:2spec}. If several filament 
are aligned in parallel
and with the same orientation, this symmetry breaking
leads to the spontaneous formation of traffic lanes for motor traffic
with opposite directionality. Varying the relative concentration of
the two motor species for $q>q_c$ leads to a discontinuous phase
transition with hysteresis, similar to the transitions induced by
changes of the external fields in magnetic systems.  In contrast to
the boundary-induced phase transitions discussed in the preceding
section, these transitions do not depend on the boundaries, but are
induced by the binding and unbinding dynamics. 
This implies that they are hardly affected by the choice of 
the boundary conditions.
%Correspondingly, they
%are robust against the choice of boundary conditions to a large
%extent. 
They depend, however, on the active movements of the motors,
and are not found in an equilibrium situation with motor velocity
$v_\bd=0$, as applies, e.g., to systems without ATP.

\section{Summary}

In summary, molecular motors exhibit interesting movements on several
length scales. Here we have addressed two of these length scales,
namely the walks along filaments which consist typically of 
$\sim$~100 steps with a step size of the order of 10nm and random
walks which consist of many such walks along filament interrupted by
periods of diffusion after unbinding from the filament. In addition,
the presence of many motors leads to traffic phenomena such as traffic
jams and traffic lanes, similar to the macroscopic traffic on streets.
However, unbinding of motors from the filaments due to thermal
fluctuations (which is a consequence of their microscopic size) plays
an important role and can help to circumvent obstacles and to
regulate the traffic.

\appendix
\section{Continuous time random walks}
\label{sec:app}

In this appendix, we summarize some results for random walks with a dwell time distribution $\psi(\tau)$ and a step
distribution $\mathcal{P}(x_s)$ which are used in section
\ref{sec:RW}. 
We consider random walkers in one dimension which make the first step at time $t=0$ starting from the origin, $x=0$.  The probability distribution $p(x,t)$ of such a random
walk fulfills the recursion relation
\cite{Montroll_Weiss1965,Weiss1994}
\begin{equation}
  p(x,t)=\sum_{n=0}^{\infty}p_n(x)\int_0^t \psi_n(t')\Psi(t-t')\d t',
\end{equation}
where $p_n(x)$ is the probability density that the walker is at
position $x$ after the $n$'th step, $\psi_n(t)$ is the probability that the
$n$'th step occurs at time $t$, and $\Psi(t)\equiv\int_t^\infty
\psi(\tau)\d\tau$ is the probability that no step occurred until time
$t$.  The initial conditions are $p_0(x)=\delta(x)$ and 
$\psi_0(t)=\delta(t)$. The solution of this recursion can be obtained using
Fourier--Laplace transforms, %see, e.g., \cite{Weiss1994}, 
which leads
to
\begin{equation}\label{CTRW_sq}
  p(q,s)=\frac{1-\psi(s)}{s[1-\mathcal{P} (q)\psi(s)]}
\end{equation}
for the Fourier--Laplace transform of the probability distribution
$p(x,t)$ \cite{Montroll_Weiss1965} with the Laplace transform of the
waiting time distribution,
%\begin{equation}
  $\psi(s)\equiv\int_{0}^{\infty} \d t\, \psi(t) e^{-st}$
%\end{equation} 
and the Fourier transform of the step distribution
%\begin{equation}
  $\mathcal{P}(q)\equiv\int_{-\infty}^{\infty}\d x_s \mathcal{P}(x_s) e^{iqx_s} \approx 1+i\langle x_s\rangle q -\langle x_s^2\rangle q^2/2$.
%\end{equation}  
The latter expansion is valid for  small $q$, provided that the moments  $\langle x_s^n\rangle$ of the step distribution $\mathcal{P}(x_s)$ with $n=1,2$ are finite. Using this expansion, 
we can derive expressions for the Laplace
transforms of the moments of our random walk by expanding $p(q,s)$ as given in \eq{CTRW_sq} in powers of $q$
\cite{Shlesinger1974} which leads to \eq{asMoments_PSIabh} from which 
the asymptotic behavior of the time-dependent moments can by obtained
via the Tauberian theorems, see \bref{Weiss1994}.

\section{Return to the origin}
\label{sec:app:return}

In this appendix we sketch the derivation of the distribution of
return times to the origin which determines the dwell time
distribution $\psi$ used in section \ref{sec:asymptotics}. We consider
the probability $f(\mathbf{x},t)$ that, at time $t$ the random walker
is at position $\mathbf{x}$ for \emph{the first time} provided it
started at the origin at time $t=0$. $f(\mathbf{0},t)$ then determines
the distribution of return times to the origin. These probabilities
satisfy the recursion relation
\begin{equation}
  p(\mathbf{x},t)=\sum_{\tau=1}^{t}f(\mathbf{x},t-\tau) p(\mathbf{0},\tau) +\delta_{t,0}\delta_{\mathbf{x},\mathbf{0}}
\end{equation}
which states that the walker is at site $\mathbf{x}$ at time $t$, if
it had been there at any time $\tau<t$ and returned there in time
$t-\tau$. The last term expresses the initial conditions.  Using
Fourier--Laplace transforms, one can derive an expression for the
Laplace transform of $f(\mathbf{0},t)$, the probability that the
walker returns to the origin at time $t$ \cite{Polya1921,Weiss1994},
\begin{equation}
  f(\mathbf{0},s)=\frac{1}{1+s}-\frac{1}{(1+s)^2 J^{(d)}(s)} %\qquad{\rm with}\qquad J^{(d)}(s)=\frac{1}{(2\pi)^d}\int_0^{2\pi}\d^d q\, p(\mathbf{q},s),
\end{equation}
where the $J^{(d)}(s)$ are integrals over the momentum $\mathbf{q}$
which depend on the spatial dimensions. The same integrals have to be
calculated for the exact solution of the motor walks
\cite{Nieuwenhuizen__Lipowsky2004}.

In the one-dimensional case we have
\begin{equation}
  J^{(1)}(s)=\frac{1}{2\pi}\int_0^{2\pi} \frac{\d q}{s+\frac{1}{2}(1-{\rm cos} q)}=\frac{1}{\sqrt{s+s^2}}
\end{equation}
which leads to $f(\mathbf{0},s)\approx 1-\sqrt{s}$ for small $s$ and
\begin{equation}
f(\mathbf{0},t)\approx \frac{1}{2\sqrt{\pi}t^{3/2}}
\end{equation}
for large $t$.  In the two-dimensional case the corresponding integral
is given by
\begin{eqnarray}
  J^{(2)}(s) & = &\frac{1}{(2\pi)^2}\int_0^{2\pi}\int_0^{2\pi}  \frac{\d q_1\,\d q_2}{s+\frac{1}{3}(2-{\rm cos} q_1-{\rm cos} q_2)}\nonumber\\
  & = & \frac{3\sqrt{m}}{\pi} K(m),
\end{eqnarray}
where $K(m)$ is a complete elliptic integral of the first kind
%\cite{Abramowitz},
%\begin{equation}
%  K(m)=\int_0^{2\pi}\frac{\d\phi}{\sqrt{1-m \sin^2\phi}},
%\end{equation}
and $m\equiv 4/(2+3s)^2$.  Since $K(m)$ behaves as
$K(m)\approx\frac{1}{2}\ln\frac{16}{1-m}$ for $m$ close to one
\cite{Abramowitz}, the return time distribution is asymptotically
given by $f(\mathbf{0},s)\approx 1-2\pi/(3\ln s^{-1})$ for small $s$
or
\begin{equation}
f(\mathbf{0},t)\approx \frac{2\pi}{3 t\ln^2 t}
\end{equation}
for large $t$.  

The dwell time distribution $\psi(\tau)$ used in
section \ref{sec:asymptotics} for the case of a filament within a
$d$-dimensional lattice is given by $\psi(\tau)=f(\mathbf{0},t=\tau)$,
where the expression on the right hand side has to be taken in the
$d_\perp=d-1$ dimensions.

%\bibliographystyle{unsrt}
%\bibliographystyle{elsart-num}
%\bibliography{../../Bibliographien/motoren,../../Bibliographien/DrivenLatticeGases,../../Bibliographien/meinePapiere,../../Bibliographien/Tools,../../Bibliographien/sonstiges,../../Bibliographien/Biology,../../Bibliographien/RandomWalks}

\end{document}